\documentstyle [12pt,a4]{article}
\begin{document}
\begin{titlepage}
\newpage
\setcounter{page}{0}
\null
\vspace{2cm}

\begin{center}
{\Large {\bf $W_{1+\infty}$, Similarity Transformation and
\vspace{0.5cm}

Interplay
Between Integer and Fractional
 \vspace{0.5cm}

Quantum Hall Effect}}
 \vspace{1cm}

{\large M.Eliashvili} $^{1,e}$
  \vspace{0.5cm}

{\it Laboratoire de Physique Th\'eorique}\/ \rm
{\small E}N{\large S}{\Large L}{\large A}P{\small P} $^\ddag $ \\
{\it Chemin de Bellevue, B.P. 110, F-74941 Annecy-le-Vieux, Cedex, France.}

\end{center}

\vspace{1cm}

\centerline{\bf Abstract}

\indent

 \rm  We  consider non-unitary similarity transformation, interconnecting   the
$W_{1+\infty}$ algebra representations  for the fractional $\nu=\frac{1}{2p+1}$
and integer $\nu=1$ filling fractions.  This transformation corresponds to the
introduction of the complex abelian Chern-Simons gauge potentials, in terms
 of which the field-theoretic description of FQHE can be developed.
The Jain's composite fermion approach and Lopez-Fradkin equivalence
assertion  are considered from the point of view of unitary and similarity
 transformations. As an application the second-quantized form of Laughlin
 function is derived.

\vspace{1cm}

\rightline{{\small E}N{\large S}{\Large L}{\large A}P{\small P}-A-466/94}
\rightline{HEP-TH/9404090}
\rightline{April 1994}
\vspace*{\fill}

$\;$
\hrulefill\ $\; \; \; \; \; \; \; \; \; \; \; \; \; \; \; \; \; \; \; \;\;$
\hspace*{3.5cm}\\
\noindent
{\footnotesize $\;\ddag$ URA 14-36 du CNRS, assosi\'e  \`a l'E.N.S. de
Lyon, et au L.A.P.P. (IN2P3-CNRS) d'Annecy-le-Vieux}

 {\footnotesize $\;^1$ On leave of absence from Tbilisi Mathematical Institute,
  Tbilisi 380093, Georgia}

{\footnotsize $\;^e$ E-mail adress: merab@lapvax.in.2p3.fr}

\end{titlepage}

$1^0$.The current understanding of the quantum Hall effect is essentially based
on the Laughlin's picture of the incompressible two dimensional quantum
fluid, which exhibits an energy gap\cite{Laugh}.

 The notion of incompressibility recently was related to the infinite symmetry,
 which
on the classical level is represented by the group of area preserving
 diffeomorphisms
 \cite{Cap1},
\cite{Cap2},
  \cite{Iso}.
 As a outcome two dimensional quantum fluid can be characterized
by the unitary irreducible highest weight representations of the $W_{1+\infty}$
algebra \cite{Cap3}.

The derivation of this basic conclusion is straightforward for the IQHE, when
liquid is formed by the noninteracting planar electrons in the lowest Landau
level $(\nu=1)$. For the clarity and to fix the notations we'll reproduce
some essential points.

In the appropriately chosen system of units $(c=\hbar=m=1,e=2,B=1)$, and
symmetric gauge ${\bf {A}}=\frac {B}{2}(-y,x)$ \/ \rm
 the quantum-mechanical Hamiltonian and angular
momentum of
N electrons in the orthogonal uniform magnetic field
 $B=\varepsilon _{\alpha \beta}\partial_\alpha A^{\beta}$ ($ {\alpha},{\beta}
=1,2$)
 can be written in terms of harmonic oscillator operators:

$$
\hat {H}=\frac {1}{2m}\sum_{k=1}^{N}[{\bf p}_k - e{\bf }A({\bf r}_k)]^2 =
\sum_{k=1}^{N}(a_k a^+_k +a^+_k a_k),
$$

$$
\hat {J}=\sum_{k=1}^{N}(z_k\partial_k -
\bar z_k\bar\partial_k)=
\sum_{k=1}^{N}[b^+_kb_k - a^+_ka_k]
$$

In the complex notations $z=x+iy, \partial =\frac{1}{2}(\partial_x -i\partial
_y)$, these operators are given by
$$
a_k=\frac{z_k}{2} +\bar \partial_k \hspace{2cm}
 a^+_k=\frac{{\bar z_k}}{2} -\partial_k
$$

$$
b_k=\frac{\bar z_k}{2} +\partial_k \hspace{2cm}  b^+_k=\frac{z_k}{2} -
\bar\partial_k
$$

$$
[a_k,a^{+}_l]=[b_k,b^+_l]=\delta_{kl}
$$
The $W_{1+\infty}$ is generated by the operators
$$
v^i_n=-\sum_{k=1}^N(b^+_k)^{n+i}(b_k)^i,\hspace{2cm} i\geq 0, \hspace{1cm}
n+i\geq 0,
$$
 which commute with the Hamiltonian and satisfy the commutation
relations
\begin{equation}
[ v^i_n,v^j_m]=(jn-im)v^{i+j-1}_{n+m}+\cdot \cdot \cdot
\end{equation}
where multidots correspond to the quantum deformations \cite{Cap1}.

The $\nu=1$ ground state is given by the wave function:
\begin{equation}
\Psi_0 (z_1,...,z_N)=\prod_{1\leq k<l\leq N}(z_k-z_l)e^{-1/2 \sum_k|z_k|^2}
\end {equation}

$$
\hat{H}\Psi_0=E_0\Psi_0= N \Psi_0
$$

$$
\hat{J}\Psi_0=\frac {N(N-1)}{2}\Psi_0
$$

The action of generators $v^i_n$ on this state can be easily
 calculated (especially if on uses
second quantization formalism). The basic results are as follows
\cite{Cap1},\cite{Cap3}:

\begin{equation}
\begin{array}{l}
a)\hspace{15pt}v^i_n\Psi_0=0\vspace{5mm}
\hspace{3,5cm}for\hspace{1cm}-i\leq n<0,i\geq1  \\

\vspace {5mm}

b)\hspace{15pt} v^i_0\Psi_0=const\cdot\Psi_0 \\

\vspace {5mm}

c)\hspace{15pt} v^i_n\Psi_0=\Phi^i_n(z_1,...,z_N)\cdot\Psi_0  \hspace{2cm} for
  \hspace{1cm} n\geq 0,i\geq1
\end{array}
\end{equation}

Here $\Phi^i_n(z_1,...,z_N)$ is some symmetric polynomial.

The equality $a)$ in (3) is  the highest weight condition,\/ \rm which is the
mathematical transcription of the incompressibility. $b)$ and $c)$ characterize
the excitation spectrum.
\vspace{5mm}

$2^0$.The situation is drastically changed in the case of fractional fillings.
Now the ground state (for $\nu=\frac{1}{2p+1}$, $p$-integer) is given by
the Laughlin wave function
\begin{equation}
\Psi_p(z_1,...,z_n)=\prod_{1\leq k<l\leq
 N}(z_k-z_l)^{2p+1}e^{-1/2\sum_k|z_k|^2}
\end{equation}
and is believed to describe the incompressible state of  interacting\/ \rm
electrons (see e.g \cite {Prange} - \cite {Stone}).

Now if one wants to construct the algebraic classification of quantum fluid,
the ground state (4) must be subjected to the action of the symmetry
generators. In order to carry out these calculations, in the recent paper
 \cite {Flo} the authors
 have changed the definition of operators $b_k$, introducing an
interaction term:
\begin{equation}
b_k\Longrightarrow B_k=b_k-2p\sum_{l\neq k}\frac{1}{z_k-z_l}
\end{equation}
Note, that $b^+_k$ is not changed
\begin{equation}
b^+_k\Longrightarrow B^+_k=b^+_k
\end{equation}

The infinite symmetry is generated by the operators
$$
V^i_n=-\sum_{k=1}^N(B^+_k)^{n+i}(B_k)^i,
$$
which satisfy the same algebra as $v^i_n$ in (1), up to the terms involving
delta-functions. These terms can be ignored, because the wave functions vanish
as $z_k \rightarrow z_l$. As a result, it can be shown that $V^i_n$ acts
on $\Psi_p$ as on the highest weight state.

The operators $B_k$ and $B^+_k$  are not Hermitian conjugate. This will be
 improved, if one introduces the new integration
measure in the configuration space, i.e
$$
dz_1 \cdot \cdot \cdot dz_N\Longrightarrow dz_1 \cdot \cdot \cdot dz_N \mu (z,
\bar{z}),
$$
where
$$
\mu(z,\bar{z})=\prod_{k< l}|z_k-z_l|^{-4p},
$$
and simultaneously changes the definition of operators $a_k$ and $a^+_k$ in the
following way:
$$
a_k\Longrightarrow A_k=a_k
 $$

$$
a^+_k \Longrightarrow A^+_k=a^+_k +2p\sum_{l \neq k}\frac{1}{z_k-z_l}
$$
(for the details and further consideration see \cite{Flo}).

Remark, that the newly introduced operators $B_k$ act on the ground state
of interacting electrons $\Psi_p$ in a way analogous to the action of
 $b_k$'s on the $\Psi_0$:
$$
b_k\Psi_0=\sum_{k\neq j}\frac{1}{z_k-z_j}\Psi_0
$$
$$
B_k\Psi_p=\sum_{k\neq j}\frac{1}{z_k-z_j}\Psi_p
$$

It seems  that this circumstance had initiated the \it Ansatz\/-\rm
 type substitutions (5-6), which in turn leads to the introduction
of the measure $\mu(z,\bar z)$ and operators $A_k$ and $A^+_k$.
\vspace{5mm}

$3^0$.In the previous note \cite {ME}
 we've made a simple observation, which perhaps clarifies the meaning of
 this procedure:  the wave functions and algebra generating operators
for the fractional $(\nu=\frac{1}{2p+1})$ and integer $(\nu=1)$ filling
fractions are related by the following {\em similarity transformation}\/  \rm:

\begin{equation}
\Psi_p(z_1,...,z_N)=S_p(z_1,...,z_N)\Psi_0(z_1,...,z_N)
\end{equation}
\begin{equation}
\hat O_p=S_p(z_1,...,z_N){\hat {O}}_0 S^{-1}_p(z_1,...,z_N),
\end{equation}
where
\begin{equation}
S_p(z_1,...,z_N)=\prod_{k<l}(z_k-z_l)^{2p}
\end{equation}
(7) is evident  ($S_p$ is a mapping operator   $T_{n-m}$
 between the ground states corresponding to the different  filling fractions
 \cite{amb}),
 and (8) can be easily verified by the direct calculations,
 letting ${\large{ \hat O}}_0=\{b_k,b^+_k
a_k,a^+_k, v^i_n\}$ and ${\hat O_p}=\{B_k,B^+_k,A_k,A^+_k,V^i_n\}$
respectively.

  Following the scheme of algebraic classification \cite{Cap3} all the
essential information about a Hall fluid for the fractional filling
is encoded in  the action of symmetry generators $V^i_n$
on the highest weight state $\Psi_p$, which due to (7)-(9) can be simply
deduced from (3):
$$
V^i_n \Psi_p=S_p\cdot v^i_n \Psi_0.
$$

In particular one automatically obtains the highest weight condition
$$
V^i_n \Psi_p=0, \hspace{2cm}  for \hspace{1cm}-i\leq n <0,i\geq 1
$$

Transformation (9) becomes singular as $z_k\rightarrow z_l$, but it acts
in the space of functions, which vanish in that limit. What seems to be
more important, is that it is not an unitary
transformation:

$$
S^{\dagger}_pS_p=\prod_{k<l}|z_k-z_l|^{4p}=\mu(z,\bar z)^{-1}
$$

The last equality is not accidental.
In the Hilbert space, where act the operators $\hat O_0$ and $\hat O^\dagger_0$
the Hermitian conjugation is defined by the scalar product:
\begin{equation}
\langle \Psi |\hat O^\dagger_0 | \Phi \rangle=
\overline{\langle \Phi |\hat O_0| \Psi \rangle}
\end{equation}
It's evident, that the operators $\hat O_p$ and $\hat O^\dagger_p=
(S^{-1}_p)^\dagger O^\dagger_0 S^\dagger_p$ are not Hermitian
conjugate in the sence of (10). Introduce the metric
operator $\hat \eta$ and define a new scalar product
\begin{equation}
\langle \Psi |\hat \eta \hat O^\dagger_p | \Phi \rangle =
\overline {\langle \Phi |\hat \eta \hat O_p| \Psi \rangle}
\end{equation}
This operator is given by
$$
\hat \eta =(S^{-1}_p)^\dagger\cdot S^{-1}_p =\mu (z,\bar z)
$$

 The transformations (7-9) interconnect the ground state vectors and
spectum generating quantum operators corresponding to two different
 physical phenomena: IQHE can be understood using a picture of
noninteracting electrons, while FQHE is essentially manifestation of
 interelectron interactions.
 On the other hand one can say, that from the point of view of
algebraic classification in the sense of \/ \rm\cite {Cap3}, \em the
IQHE and FQHE are non-unitary equivalent realizations of one and
the same underlying symmetry. \/ \rm

Similar consideration relates the $\nu=m$ states and operators to the
representation of $W_{1+\infty}$ at $\nu=\frac{m}{2mp+1}$. The corresponding
similarity transfomation is given by
\begin{equation}
S_{p,m}=\prod_{I<J} \prod_{i<j}(z^I_i-z^J_j)^{K_{I,J}}\prod_I
\prod_{i<j}(z^I_i-z^I_j)^{K_{I,I}-1}
\end{equation}
where the $m\times m$ matrix $K_{IJ}$ is defined by \cite{Fro}

$$
  K=\left|
\begin{array}{cccc}
2p+1 & 2p & ... & 2p \\
2p   &2p+1 &... & 2p \\
 .   &  .  & .  & . \\
2p  &...   & 2p &2p+1
\end{array}
\right|
$$

Interesting to note, that non-unitary similarity transformations recently
have been considered  in the context of quantum gravity, where they are
related to the temporal evolution between unstable quantum backgrounds,
indicating a deep conection between string quantum gravity and
incompressible Hall fluid \cite{Ellis}.
\vspace{5mm}

$4^0$.Note, that non-unitary transformations (8) induce
 non-canonical, complex transformations
of the phase space variables
$z_k=x_k+iy_k$ and conjugated momenta ${\bar p_k=\frac{1}{2}(p_{kx}-ip_{ky})}$:

$$
z_k\rightarrow  S_pz_kS^{-1}_p=z_k,\hspace {1cm}
\bar p_k\rightarrow S_p\bar p_k S^{-1}_p=
\bar p_k+i2p\sum_{l\neq k}\frac{1}{z_k-z_l}
$$

$$
\bar {z}_k\rightarrow S_p \bar {z}_kS^{-1}_p=\bar {z}_k, \hspace {1cm}
 p_k\rightarrow S_p p_kS^{-1}_p=p_k
$$
\vspace{5mm}

The substitutions $p\rightarrow p-\frac {e}{2}f$,
 $\bar p\rightarrow \bar p -\frac {e}{2}\bar f$
  can be inerpreted as an introduction of a   {\em complex}\/ \rm
 vector potentials

$$
f_k({\bf r}_1,...,{\bf r}_N)\equiv f_{kx} +if_{ky}=0
$$
\begin{equation}
\bar{f}_k({\bf r}_1,...,{\bf r}_N)\equiv f_{kx} - if_{ky}=-i2p\sum_{l\neq k}
\frac{1}{z_k-z_l}
\end{equation}
which depend on the positions of all N particles.

Magnetic field associated to these potentials, which acts on the k-th particle,
is given by the curl:

$$
{\cal B}_k=i({\bar{\partial}_k} {\bar f}_k -{\partial}_kf_k)=
2p \pi \sum_{k\neq l}\delta(z_k-z_l)
$$
i.e. each particle sees the $N-1$ others as vortices carring a  $2p$ elementary
flux quanta. This fact has been already noted in \cite{Flo}.

 In the chosen system of units flux quantum $\phi_0=\pi$,
and the density of Landau states $n_B=\frac{eB}{2\pi}=1/\pi$. Hence,
the filling fraction

$$
\nu=\frac{N}{(\Phi/\phi_0)}=\frac{1}{2p+1},
$$
where the total flux $\Phi =\pi N(2p+1)$.

Using the mean-field arguments, one can say that electrons move
in the average magnetic field $2p+1$, in accordance with the Jain's
hierarchical construction \cite{jain}. However  the additional
magnetic field ${\cal B}=2p$ is generated now by the complex gauge potentials,
\/\rm in contrast to the composite fermion approach \cite{jain},
 where magnetic fluxes
attached to the point particles are produced by the real singular
vector potentials:
$$
{\vec{\cal A}}_k=p{\vec {\nabla}}_k \sum_{l\not= k}\Theta_{kl}, \hspace{1cm}
\Theta_{kl}=-i\arg (z_k-z_l),
$$
or in the complex notations

\begin{equation}
{\cal A} _k=ip\sum_{l\not= k}\frac{1}{\bar z_k-\bar z_l}, \hspace{10mm}
\bar {\cal A}_k=-ip\sum_{l\not= k}\frac{1}{ z_k- z_l}
\end{equation}
 The corresponding ground state is known
to be \cite{jain},\cite{laugh2}
\begin{equation}
\Phi_p=\prod_{k<l}\frac{(z_k-z_l)^{2p}}{|z_k-z_l|^{2p}}\Psi_0 (z_1,...,z_N)
\equiv U_p({\bf r}_1,...,{\bf r}_N)\Psi_0 (z_1,...,z_N)
\end{equation}
which containes one particle states from
the higher Landau orbitals and its energy is higher than the FQHE
ground state energy.

Note, that gauge potentials (14) can be introduced as a  singular
unitary transformation:
$$
U_p p_kU^\dagger_p=p_k-\frac {e}{2}{\cal A}_k
$$
$$
U_p \bar p_kU^\dagger_p=\bar p_k-\frac {e}{2}\bar {\cal A}_k
$$

The last remark can be related to the equivalence between  the system
of electrons bounded to the even number of magnetic flux quanta
 and the  same system  without  these fluxes  \cite{lopez}.
{}From our consideration it foollows, that these two theories can
be related by the unitary operator $U_p$ as well as by a similarity
transformation $S_p$. In the former case equivalence assertion
given in \cite{lopez} is in fact a quantum-mechanical unitary
equivalence.  The same time non-unitary character of $S_p$ is a
 loophole, which enables to evade the consequences of the
equivalence statement, and reduces the study of FQHE of the mutually
interacting electrons to the IQHE of nonintreacting composite
particles.
\vspace{5mm}

$5^0$. The potentials (13) and (14) have form
typical for the statistical interaction with a parameter $\theta =4p\pi$ (see
e.g. \cite{Lerda}) and naturally can be
incorporated  into the framework of the
Chern-Simons theories.

  Introduce the
 particle density at the point ${\bf r}$:

\begin{equation}
\varrho({\bf r})=\sum_{l=1}^N\delta({\bf r}-{\bf r}_l)
\end{equation}
and vector potentials  satisfying the equation:

\begin{equation}
\varepsilon_{\alpha \beta}\partial_{\alpha}a^\beta({\bf r})=2p\pi \varrho({\bf
r})
\end{equation}

The solutions to (17) can be easily found:

\begin{equation}
{\cal A}^\alpha({\bf r})=-2p\pi\varepsilon_{\alpha \beta}
\partial_{\beta}\int d{\bf r'}G({\bf r}-{\bf r'})\varrho ({\bf r'}) =
p\partial_\alpha \int d{\bf r}'\Theta (z-z')\varrho ({\bf r}')
\end {equation}
and

\begin{equation}
f^\alpha({\bf r})=-2p\pi(\varepsilon_{\alpha \beta}+i\delta_{\alpha \beta})
\partial_{\beta}\int d{\bf r'}G({\bf r}-{\bf r'})\varrho ({\bf r'})=
-ip\partial_\alpha \int d{\bf r}'\ln (z-z')\varrho ({\bf r}')
\end {equation}
\vspace{5mm}

Here $ G({\bf r})=\frac{1}{2\pi}\ln r $  is a Green function, and
$ \Theta (z)=-i\arg z $.

Substituting into (18) and (19) the particle density  (16) and letting ${\bf
r}=
{\bf r}_k$, we
immediately recover (13) and (14).

The same time (17) is a field equation for the Chern-Simons Lagrangian

 \begin{equation}
{\cal L}=i\psi^\dagger(\partial_0+i+iea_0)\psi -
\frac {1}{2} (D_\alpha \psi^\dagger)(D_\alpha \psi) -
\frac {e^2}{8\pi p} {\varepsilon}^{\mu \nu \lambda}a_\mu \partial_\nu a_\lambda
\end{equation}
$$
 D_\alpha \psi=(\partial_\alpha +ieA_\alpha +iea_\alpha)\psi ,\hspace{1cm}
  D_\alpha \psi^\dagger=(\partial_\alpha -ieA_\alpha -iea_\alpha)\psi^\dagger
$$
The solutions given by (18) and (19) produce   magnetic
field
$$
b({\bf r})=2p\pi\sum_{l=1}^N\delta({\bf r}-{\bf r}_l)
$$
 and are related by the complex gauge transformation
$$
f^\alpha ({\bf r})={\cal A}^\alpha({\bf r})-ip\partial_\alpha
\int d{\bf r}' \ln |z-z'| \varrho ({\bf r}')
$$

  The composite fermions bounded to $2p\phi_0$ magnetic fluxes
can be introduced as
a gauge transformation of real electrons. In the case of a Jain's particles
it will be \cite {Halp1}
\begin{equation}
\psi_c ({\bf r})= e^{2ip\int d{\bf r}'\Theta (z-z')\varrho (t,{\bf r}')}
\psi ({\bf r})
\end{equation}

In the case of complex gauge potential quasiparticle fields are defined
as follows:
$$
\chi_c (t,{\bf r})=
e^{-2p\int d{\bf r}'\ln (z-z')\varrho (t,{\bf r}')}\chi (t,{\bf r})
$$

\begin{equation}
\chi^\star_c (t,{\bf r})=\psi^\dagger_c (t,{\bf r})e^{2p\int d{\bf r}'\ln
(z-z')\varrho (t,{\bf r}')}
\end{equation}

Note, that $\chi^\star_c\not =\chi^\dagger_c$. The density operator
$$
\varrho (x)=\psi^\dagger (x)\psi (x)=\chi^\star_c (x) \chi_c (x)
$$

Vacuum

$$
\psi (x) |0\rangle = \langle 0|\psi^\dagger (x)=0
$$
is the same for quasiparticle field:
$$
\chi_c (x) |0\rangle = \langle 0|\chi^\star_c (x)=0
$$

  The composite particles obey fermionic statistics:
$$
\chi_c ({\bf r}_1)\chi_c ({\bf r}_2)= (-1)e^{\omega (z_2-z_1)-\omega (z_1-z_2)}
\chi_c ({\bf r}_2)\chi_c ({\bf r}_1) =
(-1)e^{-2ip\pi} \chi_c ({\bf r}_2)\chi_c ({\bf r}_1)
$$

$$
\chi_c ({\bf r}_1)\chi^\star_c ({\bf r}_2) = \delta ({\bf r}_1-{\bf r}_2)
-e^{\omega (z_1-z_2)-\omega (z_2-z_1)}
\chi^\star_c ({\bf r}_2)\chi_c ({\bf r}_1)
$$

$$
= \delta ({\bf r}_1-{\bf r}_2)
-e^{2ip\pi}\chi^\star_c ({\bf r}_2)\chi_c ({\bf r}_1)
$$

Here

$$
\omega(z_1-z_2)=-2p\ln (z_1-z_2)
$$

The matter Hamiltonian
\begin{equation}
H=\int d{\bf r}\psi^\dagger ({\bf r})[-\frac {1}{2}D_\alpha D_\alpha -
ef_0({\bf r})]\psi ({\bf r})
\end{equation}
can be expressed in terms of composite particle fields
\begin{equation}
H=\int d{\bf r}\chi^\star_c ({\bf r})[-\frac {1}{2}\nabla_\alpha \nabla_\alpha
]\chi_c ({\bf r})
\end{equation}

where
$$
D_\alpha=\partial_\alpha +ieA_\alpha +ie f_\alpha
$$
and

 $$
\nabla_\alpha=\partial_\alpha +ieA_\alpha
$$

Calculate
$$
\langle z_1,...,z_N|=\langle 0|\psi ({\bf r}_1) \psi (z_2) \cdot \cdot
\cdot \psi ({\bf r}_N)=  \vspace{5mm}     \\
$$
$$
= \langle 0 |e^{2p\int d{\bf r}' [\sum_{k=1}^N \ln (z-z_k)]\varrho ({\bf r}')}
\prod_{j< l}e^{2p\ln (z_j-z_l)} \chi_c ({\bf r}_1)
\chi_c ({\bf r}_2)\cdot \cdot\cdot \chi ({\bf r}_N)=\vspace{5mm}   \\
$$
$$
=\prod_{k< l}(z_k-z_l)^{2p}
\langle 0|\chi_c ({\bf r}_1)\cdot \cdot \cdot \chi_c ({\bf r}_N)
$$
\vspace{5mm}

Expand $\chi_c ({\bf r}) $  into modes
$$
\chi_c ({\bf r})=\sum_a f_a u_a({\bf r})
$$
and
$$
\chi^\star_c({\bf r})=\sum_a f^+_a \bar {u}_a({\bf r})
$$
where
$$
\lbrace f_a,f^+_b \rbrace =\delta _{ab}
$$
Let
$$
|\Psi \rangle = f^+_{a_1}\cdot \cdot \cdot
f^+_{a_N} |0 \rangle
$$
be en eigenstate of hamiltonian $H$. Then

$$
\langle z_1,...,z_N|\Psi\rangle = \prod_{k\not= l}(z_k-z_l)^{2p}
\langle 0|\chi_c ({\bf r}_1)\cdot \cdot
\cdot \chi_c ({\bf r}_N)f^+_{a_1}\cdot \cdot \cdot f^+_{a_N} |0\rangle
$$
where last factor is a Slater determinant $ \Psi_0 (z_1,...,z_N) $.
Hence the Laughlin wave function can be considered as a coordinate
space representation of the composite fermion ground state.

Concluding, we can say the following:  According to \cite{Cap3}  the quantum
states of the incompressible fluid can be exhaustively classified
by the unitary irreducible
highest weight representations of the algebra $W_{1+\infty}$. Applying to the
representation at $\nu=$integer the similarity transformation (8) or (10)
 one
 automatically (at least in principle)
 obtains the corresponding classification for the fractional values
of filling fraction. This transformation seems to be equivalent to the
introduction of the complex abelian C-S gauge potentials in terms of which
the field-theoretic description of FQHE can be given.

\vspace{3cm}
{\large {\bf Acknowlegments}}\/\rm \vspace{0.5cm}

I'm particularly thankful to P.Sorba for bringing the considered problems
to my attention, helpful discussions and support. It is a pleasure to
thank N.Mavromatos for a useful discussion.

\vspace{0,5cm}

  \end{document}